\documentclass[twocolumn,showpacs,preprintnumbers,nofootinbib,prd,superscriptaddress,groupedaddress,10pt]{revtex4-1}

\usepackage{amsmath}
\usepackage{graphicx,amssymb,amsthm,amsfonts,epsfig,epsf}
\usepackage[linktocpage]{hyperref}
\usepackage[usenames]{color}
\usepackage{epstopdf}
\usepackage{amsmath}
\usepackage{dcolumn}
\usepackage[latin1]{inputenc}
\usepackage{latexsym}
\usepackage{rotating}
\usepackage{hyperref}
\usepackage{color}
\usepackage{longtable}
\usepackage{enumerate}
\usepackage{tensor}
\usepackage{mathtools}
\usepackage{natbib}
\usepackage{booktabs}
\usepackage{url}
\def\be{\begin{equation}}
\def\ee{\end{equation}}
\def\bea{\begin{eqnarray}}
\def\eea{\end{eqnarray}}
\def\ben{\begin{equation*}}
\def\een{\end{equation*}}
\newcommand{\beq}{\begin{eqnarray}}
\newcommand{\eeq}{\end{eqnarray}} 
\newcommand{\ba}{\begin{align}}
\newcommand{\ea}{\end{align}}

\def\nn{\nonumber}

\begin{document}

\title{Boson Stars in Higher Derivative Gravity}
%
\author{
Vishal Baibhav
}
\email{baibhavv@gmail.com}

\author{
Debaprasad Maity
}
\email{debu.imsc@gmail.com}
\affiliation{ Department of Physics, Indian Institute of Technology, Guwahati, India}

\begin{abstract}
In this paper, we have constructed Boson star (BS) solutions in four dimensional scalar-Gauss-Bonnet (sGB) theory.
In order to have non-trivial effect from Gauss-Bonnet term, we invoked non-minimal coupling 
between a complex scalar field and the Gauss-Bonnet term with a coupling parameter, $\alpha$.  We show that the scalar field can no longer take arbitrary value at the center of the star. 
Furthermore, boson-stars in our higher derivative theory turn out to be slightly massive but much more 
compact than those in the usual Einstein's gravity. Interestingly, we found that for $\alpha<-0.4$ and $\alpha>0.8$, binding energy for all possible boson stars is always negative. This implies that these stars are intrinsically stable against the decay by dispersion. We also present the mass-radius and mass-frequency 
curves for boson-star and compare them with other compact objects in gravity models derived from Gauss-Bonnet term.
\end{abstract}
\maketitle

\section{Introduction}
Boson stars are self gravitating and localized solutions of a complex scalar field. These exotic stars are held back from collapsing under their own self gravity by Heisenberg's uncertainty principle. This is in contrast with compact objects like neutron stars where degeneracy pressure (from Pauli's exclusion principle) prevents the gravitational collapse. Despite having no observational evidence, these  star like configurations are widely studied as black-hole mimickers, dark matter candidates and often as astrophysical compact objects. 
If constructed without any coupling to electromagnetic field, they could serve as horizon-less black holes, or if constructed on galactic scales they might act as dark-matter halo. Furthermore, boson-stars share a large number of similarities with neutron stars: both form one parameter family, have similar mass-radius curves and show transition from stable to unstable states at the peak of these curves. Because of such similarities, boson stars are often studied as astrophysical compact objects.
Even if they are found non-existent in
nature, they could serve as analogs to neutron stars which are sometimes harder to model (See \cite{Liebling:2012fv,Jetzer:1991jr} for a review). Because of their sheer simplicity, they make perfect tools to explore General Relativity.\\

In order to completely understand the nature of a classical solution such as that of a  boson star, 
it is necessary to probe the dynamics of small perturbations around aforementioned 
classical solution. Studying the effects of an higher dimensional operator,
which can be thought of as one such perturbation at the effective Lagrangian level, can be an important way that can shed 
light not only into the nature of a boson star but also into the nature of the effective field theory itself. In the present paper, we will 
take into account a specific form of higher derivative term in the gravity sector and study its static modification
on usual minimal boson star solutions.
There exits an interesting generalization of General Relativity to an arbitrary number of dimensions, known as Lovelock's theory,
which contains an action comprising of higher order derivatives. One of the most interesting properties of this higher derivative theory is that it does not 
lead to more than two derivatives at the equation of motion level. Theories with more than two derivatives are, in general, plagued by ghosts and consequently their Hamiltonians are unbounded from below. The term quadratic in curvature in Lovelock's generalization is called the Gauss Bonnet term. Boson stars have been extensively studied in Gauss Bonnet geometry in five dimensions \cite{BH,SKV,MBS5D,RBS5D}. However in four dimensions, the Gauss Bonnet term is a topological invariant and gives no contribution to the equations of motion. This problem can be avoided by introducing a non-minimal 
coupling between a complex scalar field and the Gauss Bonnet term. In order to have global phase rotation invariance,
Gauss-Bonnet modification should couple with the square of the complex scalar field. 
Boson stars with the non-minimal scalar field coupled to curvature term has been studied earlier by \cite{bij,bsnm}, where they coupled Ricci scalar to square of boson field. Therefore, it would be a simple extension to study the influence of coupling to higher curvature terms.\\
In this work, we will extensively study ground-state boson stars in a scalar-Gauss-Bonnet (sGB) theory. In addition to standard BS Lagrangian, we have a Gauss-Bonnet term non-minimally coupled to the complex bosonic fields. For simplicity,
we have not considered any self interaction term for scalar field in the Lagrangian. However, it 
would be straight forward to extend our analysis if one includes quartic coupling in the scalar sector.
 Boson stars without self interactions, often called mini-boson stars, have maximum mass of the order $M_{pl}^2/m$.
This is much smaller than the Chandrasekhar mass $M_{pl}^3/m^2$ for the fermionic counterparts, where $m$ is
the mass of the field under consideration. To construct boson stars with larger masses and more particles, 
a repulsive self-interaction term is needed to provide the extra pressure against gravitational collapse. Even though we consider mini-boson stars in this paper, as mentioned the analysis can be easily extended to include the self interaction term needed to reach the astrophysical mass scales. Furthermore, we will mainly concentrate on the solutions with positive coupling constants, for reasons that will be clear as we go along.\\
 The paper is organized as follows. In section \ref{sec:model} we present the underlying Lagrangian formulation and equations of motion. Boundary conditions are discussed and it is shown that the parameter space is constrained.
 In Sec \ref{sec:analysis}, we discuss different properties of the numerical solutions like binding energy, compactness, M-R and M-$\omega$ curves.

\section{Model and Setup \label{sec:model}}
We start with an action containing the standard boson star Lagrangian and the Gauss Bonnet term non-minimally coupled to 
the product of bosonic fields $\phi \text{ and } \phi^*$.

\bea
S=\int  d^4x \sqrt{-g}\left(\frac{R}{16 \pi  G}+\mathcal{L}_\phi+\mathcal{L}_\text{GB}\right) \nn \\
\mathcal{L}_\phi=-g^{\mu  \nu } \nabla _{\mu }\phi  \nabla _{\nu }\phi ^*  -m^2 \left| \phi \right| ^2 \nn\\
\mathcal{L}_\text{GB}=\frac{A}{4 M_\text{pl}^2}\phi \phi ^* \mathcal{G}
\label{eq:Lagrangian}
\eea
Here $\mathcal{G}=R^2+R_{\mu \nu \rho \sigma } R^{\mu \nu \rho \sigma }-4 R_{\mu \nu } R^{\mu \nu }$ is the unique combination of Riemann curvature terms that retains the second order nature of field equations, A is dimensionless coupling parameter and m is the mass of boson. Varying the above Lagrangian with conjugate scalar $\phi^*$ and  $g_{\mu\nu}$  yields Klein-Gordon and Einstein's equations respectively

\bea
\Box \phi=m^2\phi-\frac{A}{4 M_\text{pl}^2} \mathcal{G}\phi\\
R_{\mu\nu}-\frac{1}{2}g_{\mu\nu}R=8 \pi G (T_{\mu\nu}^{\phi}+T_{\mu\nu}^{GB})
\eea

where contribution to stress-energy tensor by bosonic matter is given by
\begin{multline}
T_{\mu\nu}^{\phi}=\nabla _{\nu }\phi  \nabla _{\mu }\phi ^*+\nabla _{\mu }\phi  \nabla _{\nu }\phi ^*-g_{\mu \nu }m^2 \left| \phi \right| ^2\\ -g_{\mu \nu } (g^{ab} \nabla _b\phi  \nabla _a\phi ^*)
\end{multline}
Presence of Gauss Bonnet term modifies Einstein's equation with addition of term $T_{\mu\nu}^{GB}$ \cite{DE,Nojiri:2007te}
\begin{multline}
T_{\mu\nu}^{GB}=(\nabla _{\mu }\nabla _{\nu }F)R-g_{\mu \nu } \left(\nabla _{\rho }\nabla ^{\rho } F\right)R-2\left(\nabla ^{\rho } \nabla _{\mu }F\right)R_{\nu \rho }\\
-2\left(\nabla ^{\rho }\nabla _{\nu }F\right)R_{\mu \rho }+2\left(\nabla _{\rho }\nabla ^{\rho }F \right)R_{\mu \nu } \\
+2g_{\mu \nu } \left(\nabla ^{\rho }\nabla ^{\sigma }F\right)R_{\rho \sigma }-2\left(\nabla ^{\sigma }\nabla ^{\rho }F \right)R_{\mu \rho \nu \sigma }
\end{multline}

Considering harmonic ansatz for the scalar field $\phi(t,r)=\phi _0(r) e^{-i \omega t}$, and a static, spherically-symmetric space-time metric with line element 
\begin{eqnarray}
ds^2=-e^{\nu} dt^2+e^{\lambda}dr^2 +r^2 \left(d\theta ^2+\sin^2\theta d\varphi ^2\right)\,
\end{eqnarray}
we get a resulting system of three coupled equations that needs to be solved numerically for scalar field amplitude $\phi_0(r)$ and two metric fields $\lambda(r),\nu(r)$. 

\begin{widetext}

\begin{multline}
\frac{e^{\nu -\lambda } \left(e^{\lambda }+r \lambda '-1\right)}{8 \pi G r^2}=
e^{\nu -\lambda } \phi _0'^2+\phi _0^2 e^{\nu } f +
\frac{A}{M_{pl}^2}\frac{2 e^{\nu -2 \lambda } \left(\phi _0 \left(2 \left(e^{\lambda }-1\right) \phi _0''-\left(e^{\lambda }-3\right) \lambda ' \phi _0'\right)+2 \left(e^{\lambda }-1\right) \phi _0'^2\right)}{r^2}\nn
\end{multline}

\bea
\frac{-e^{\lambda }+r \nu '+1}{8 \pi G r^2}=
e^{\lambda } \phi _0^2 g+\phi _0'^2
-\frac{A}{M_{pl}^2}\frac{2  \left(1-3 e^{-\lambda }\right) \phi _0 \nu ' \phi _0'}{r^2}\nn\\
e^{-\lambda } \phi _0''+e^{-\lambda } (\frac{\nu '- \lambda '}{2}+\frac{2}{r}) \phi _0'+g
\phi _0=
-\frac{A}{M_{pl}^2}\frac{\left(e^{\lambda }-3\right) \lambda ' \nu '-\left(e^{\lambda }-1\right) \left(2 \nu ''+\nu '^2\right)}{2 r^2}e^{-2 \lambda } \phi _0 \label{eq:eom}
\eea
\end{widetext}
where  $f=\left(m^2+\omega ^2 e^{-\nu }\right)$ and $g=\left(e^{-\nu } \omega ^2-m^2\right)$.

During numerical integration, we use the following  dimensionless variables
\begin{eqnarray*}
\tilde{\omega}=\omega/m\text{ , }\tilde{r}=r m \text{ , }\sigma = \phi_0 \sqrt{8\pi G}\text{ and }\alpha=A \frac{m^2}{M_\text{pl}^2}
\end{eqnarray*}\\ 
\\

In this paper, we will be studying the following physical quantities characterizing a boson star. \\ 
{\it Mass}(M): Asymptotically any boson star metric resembles Schwarzschild metric, which allows us to associate the metric
coefficient $e^{\lambda(r)}=(1- 2M/r)^{-1}$, where M is the ADM mass defined for an asymptotically flat space-time. It can be calculated as
\be
M=\frac{r_{\text{out}}}{2}(1-e^{-\lambda(r_{\text{out}})})
\ee
where $r_{\text{out}}$ is the outermost point of numerical domain.\\
{\it Particle Number}(N): Boson star given by \ref{eq:Lagrangian} is globally $U(1)$ invariant.
In other words, the given system is invariant under global phase rotation implying a locally conserved 
Noether current $J^{\mu }=\frac{1}{2} i g^{\mu \nu } \left(\phi ^* \nabla _{\nu }\phi -\phi  \nabla _{\nu }\phi ^*\right)$. 
The corresponding U(1) charge of the boson star gives the total number of bosons (N) in the star under consideration.
\be
N=\int  \sqrt{-g} J^0 d^3x=4 \pi \int r^2 \omega  \phi _0^2 e^{\frac{\lambda-\nu }{2}}
\ee
\\
{\it Binding Energy}$(E_b)$: Binding energy $E_b=M - N m$ plays a crucial role in determining the classical gravitational stability of an astrophysical compact object. A positive binding energy implies an unstable configuration. Particles in such systems have more kinetic energy than the gravitational energy holding them together. Consequently they disperse to infinity. However negative $E_b$ does not guarantee stability as some configurations might still collapse into black holes. In other words, $M<mN$ is a necessary but not sufficient condition for stability. Note that solutions with negative binding energies are said to be "classically stable". We will discuss
this classical stability in subsequent sections. However, it is imperative to study the perturbation around the boson star solution to understand the
stability properties better. We will defer this study to a future work. \\
{\it Radius and Compactness}$(R_{99}, C)$: Boson stars lack a defined
surface, as theoretically they are infinitely extended objects. However, from the physical point of view, we can
always define an effective radius($R_{99}$) within which the boson star under consideration contains 99\% of the total mass($M_{99}$).
Therefore corresponding effective compactness can be defined as $C=M_{99}/R_{99}$.\\
After stating the boundary conditions, in the following section, we will try
to solve the equations of motion(\ref{eq:eom}) numerically and compute all the aforementioned physical quantities. 
\subsection{Boundary Conditions and constraints on parameter space \label{sssec:constraints}}
In absence of a Gauss Bonnet term, boson stars are parameterized by the scalar field value at the origin $\phi_0(0)$ which attain any arbitrarily large value. In presence of Gauss-Bonnet term, however, solutions exist only up to a maximum value of $\phi_0(0)$. Given a value of coupling constant A, there exists a limiting value of $\phi_0(0)=\phi_{\text{cr}}$, after which solutions do not exist. Similar constraints on the parameter space have been found for boson star in other theories, such as five dimensional  Einstein-Gauss-Bonnet(EGB) gravity \cite{SKV,MBS5D} and fluid stars in Einstein-Dilaton-Gauss-Bonnet(EDGB) gravity \cite{Pani}. This criticality appears from the regularity condition of the scalar and metric fields at origin due to higher derivative 
gravity term. To illustrate this further, we expand the metric and scalar field near the origin as follows,

\bea
\lambda (r)=\lambda_2 r^2+O(r^4)\\
 \nu (r)=\nu_0+\nu_2 r^2 +O(r^4)\\
\phi_0(r)=p_0+ p_2r^2+O(r^4)
\eea 
Regularity conditions demand that all first order derivatives and $\lambda(0)$ vanish at origin. The (tt), (rr) components of Einstein's equations and Klein Gordon equation at zeroth order now read

\bea
-24 A \lambda _2 e^{\nu _0} p_2 p_0+3 \lambda _2 e^{\nu _0}-e^{\nu _0} m^2 p_0^2-p_0^2 \omega ^2=0\label{eq:eo1}\\
-16 A \nu _2 p_2 p_0-\lambda _2+m^2 p_0^2+2 \nu _2-e^{-\nu _0} p_0^2 \omega ^2=0\label{eq:eo2}\\
-6 A \lambda _2 \nu _2 p_0-m^2 p_0+e^{-\nu _0} p_0 \omega ^2+6 p_2=0\label{eq:eo3}
\eea
These equations can be solved to obtain $\nu_2$, $\lambda_2$ and $p_2$, while $p_0$ serves as a free parameter and $\nu_0$ has to be set in such a way that $\nu(r)$ vanishes asymptotically. After eliminating $\nu_2$ and $p_2$ using \ref{eq:eo1} and \ref{eq:eo3}, one gets the quartic equation for $\lambda_2$ from \ref{eq:eo2}

\begin{multline}
-\frac{f_0^2 p_0^2}{36 A^2}+\frac{\lambda _2 \left(3 f_0+4 A f_0 g_0  p_0^2 e^{\nu _0} m^2 \right)}{36 A^2}\\-g_0 \lambda _2^3 p_0^2-\lambda _2^4
=0\label{eq:qe}
\end{multline}
where $f_0= m^2+\omega ^2 e^{-\nu _0}$ and $g_0= \omega ^2 e^{-\nu _0}-m^2$. Nature of solutions of a quartic equation can be determined by its discriminant. For a quartic polynomial of type $a \lambda _2^4+b \lambda _2^3+d \lambda _2+e$ (no quadratic term), the discriminant is given by

\begin{multline}
\Delta =256 a^3 e^3-192 a^2 b d e^2-27 a^2 d^4-6 a b^2 d^2 e\\-27 b^4 e^2-4 b^3 d^3 \label{eq:discriminant}
\end{multline}
For a generic quartic case, $\Delta<0$ means two complex-two real roots and  for $\Delta>0$, all roots are either real or complex.  Equation  \ref{eq:qe} always has two complex zeros in the regime of interest. Consequently, physical solutions exist only for negative discriminant. And as a result, the possible combinations of Gauss-Bonnet parameter, A and central value of scalar $\phi_0(0)$, are now constrained. As we will see, this imposes an upper bound on $\phi_0(0)$ for a given coupling constant.\\
Solving \ref{eq:eo1}, \ref{eq:eo2}, \ref{eq:eo3} for $p_2$, $\lambda_2$ and $\nu_2$ and accepting only physical solutions, we get the boundary conditions at "$r_0$", the starting point of numerical domain

\bea
\lambda (r_o)=\lambda_2 r_o^2\nn\\
 \nu (r_o)=\nu_0+\nu_2 r_o^2\nn\\
\phi_0(r_o)=p_0+ p_2r_o^2 \label{eq:bc}
\eea
As stated earlier, $p_0$ is the free parameter, while $\nu_0$ has to be set such that $\nu(\infty)\to0$\\
Besides regularity at origin and asymptotic flatness, we also require an asymptotically vanishing energy density. This can be attained by shooting for a certain value of $\omega$ such that $\phi_0(\infty) \to 0$. Requirement of asymptotic flatness, eliminates any role  of the Gauss-Bonnet term at the infinity as the term vanishes in flat space-time.
 Therefore, in the asymptotically flat space-time limit ($\lambda \approx 0$ and $\nu \approx0$ as $r\rightarrow\infty$), scalar field equation turns out to be
\bea
 \phi_0{''} + \frac 2 r  \phi_0' -(m^2-\omega^2) \phi_0 =0 \nonumber .
\eea

Making sure that both metric fields and amplitude of scalar field vanish at large radial distances, one gets the usual behavior of the scalar field decaying as 
$\phi \propto e^{-r \sqrt{m^2-\omega ^2}}/r$  from the 
above equation for the scalar field.

\section{Analysis  of Numerical Solutions \label{sec:analysis}}

\begin{figure*}[t]
\begin{center}
\begin{tabular}{lr}
\epsfig{file=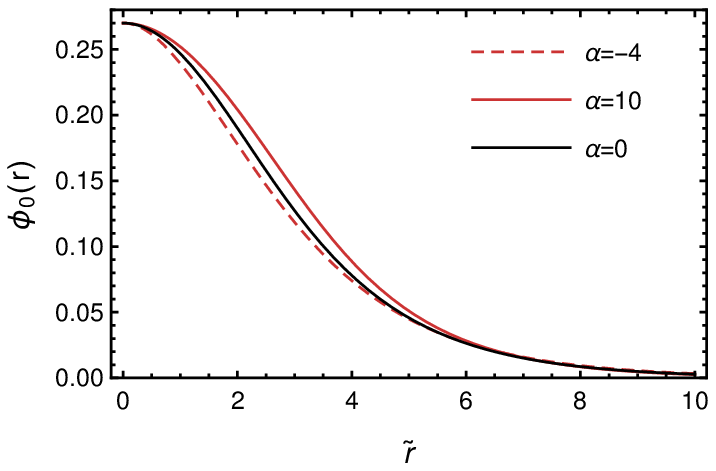, width=0.45 \linewidth}&
\epsfig{file=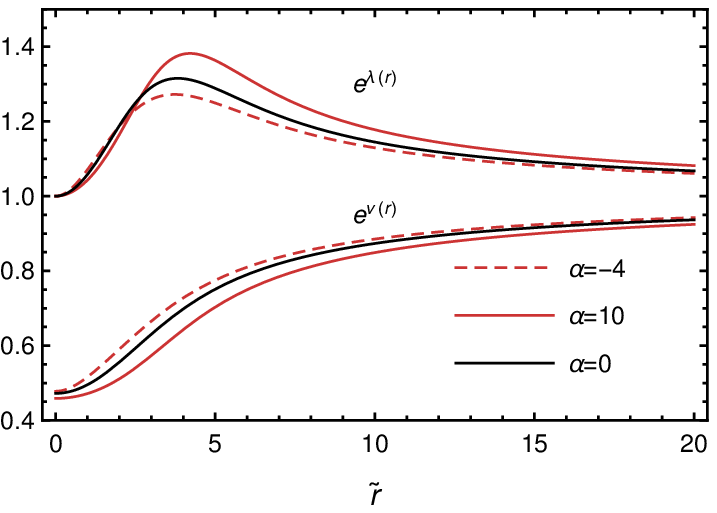, width=0.45 \linewidth}
\end{tabular}
\end{center}
\caption{\label{fig:fp} Profiles of scalar and metric fields in sGB gravity: $\phi_0(0)=0.27$
}
\end{figure*}
In this section, we'll discuss the solutions of our master equation \ref{eq:eom} given the boundary conditions \ref{eq:bc}
for scalar field amplitude $\phi_0(r)$ and two metric fields $\lambda(r),\nu(r)$ discussed in the previous section.
As shown in Fig \ref{fig:fp}, we observe a very marginal change in the appearance of scalar field profile, 
due to Gauss-Bonnet term. As stated earlier, Gauss-Bonnet term vanishes as 
we asymptotically approach the flat space-time. As a result, the scalar field equation at infinity in sGB gravity remains the same as in Einstein's gravity. This is reflected in Figure \ref{fig:fp}, where scalar fields for different coupling have same values at large radial coordinate. Effective radius $R_{99}$ reduces only slightly in our model as was predicted for usual 
Einstein's gravity with a massive complex scalar field.
We found that masses and number of bosons in boson stars for positive coupling to sGB gravity are slightly larger than Einstein boson-stars, but of same order. Consequently sGB boson stars are much more compact than those found in Einstein's gravity, as we shall discuss 
in the subsequent sections. This is in contrast with neutron stars studied by \cite{Pani} in Einstein-Dilaton-Gauss-Bonnet gravity. They studied Lagrangian for fluid stars with a real scalar field coupled  to Gauss Bonnet term and found that stars in EDGB are less massive than those in standard Einstein gravity. Even though the case studied by \cite{Pani} is entirely different from our case, we still think it's worthwhile to make comparisons whenever we can.\\
As can be seen in Fig \ref{fig:be}-\ref{fig:compact}, for low $\phi_0(0)$, both Einstein and sGB theories give same values for all physical parameters. This feature is shared by EGB boson-stars in five dimensions and \cite{MBS5D} attribute this behavior to smallness of energy momentum tensor. For scalar fields with low amplitudes, the effective Gauss-Bonnet coupling is not strong enough. As a consequence, we can't differentiate between a boson star in Einstein-Hilbert gravity or sGB gravity, if the boson star is very small.\\
We now study one by one different properties of the sGB boson-stars.

\begin{table*}
\caption{Numerical results are presented for some values of the coupling constant. The $\phi_{cr}$  and corresponding discriminant after which solutions cease to exist are given with error $10^{-5}$. Discriminant $\Delta$ is normalized so that maximum of $\left|\Delta\right|$ lies at +1, with  $\Delta_N=\frac{\Delta}{Max(\left|\Delta\right|)}$ . Closer the value of $\Delta_N$ is to zero, more accurate the $\phi_{cr}$ is. Maximum value of mass, particle number and compactness are also presented in the table. }
\label{tab:data}
\begin{tabular}{||c|c|c|c|c|c||}
\hline
$\alpha$ & $\phi_{cr}$ & $\Delta_N(\phi_{c})$ & $M_\text{max}\times (m/M_{pl}^2)$& $N_\text{max}\times (m^2/M_{pl}^2)$& $C_\text{max}/M_{pl}^2$\\

\hline
0 & $\infty$  & Not Defined              & 0.6330 & 0.6530 & 0.1109 \\
0.1 & 0.96732 & $-4.82 \times 10^{-4}$ & 0.6344 & 0.6546 & 0.1240 \\
1 & 0.80421 & $-1.04 \times 10^{-4}$ & 0.6534 & 0.6766 & 0.2061 \\
2 & 0.75679 & $-6.14 \times 10^{-4}$ & 0.7469 & 0.8012 & 0.2244 \\
3 & 0.73017 & $-4.29 \times 10^{-4}$ & 0.8184 & 0.8986 & 0.2351 \\
4 & 0.71181 & $-1.92\times 10^{-4}$ & 0.8730 & 0.9732 & 0.2404 \\
5 & 0.69786 & $-1.63\times 10^{-4}$ & 0.9172 & 1.0334 & 0.2428 \\
10 & 0.65598 & $-5.33\times 10^{-4}$ & 1.0622 & 1.2277 & 0.2521 \\
 \hline 
\end{tabular}
\end{table*}

\subsection{Parameter constraints}

\begin{figure*}[t]
\begin{center}
\begin{tabular}{lr}
\epsfig{file=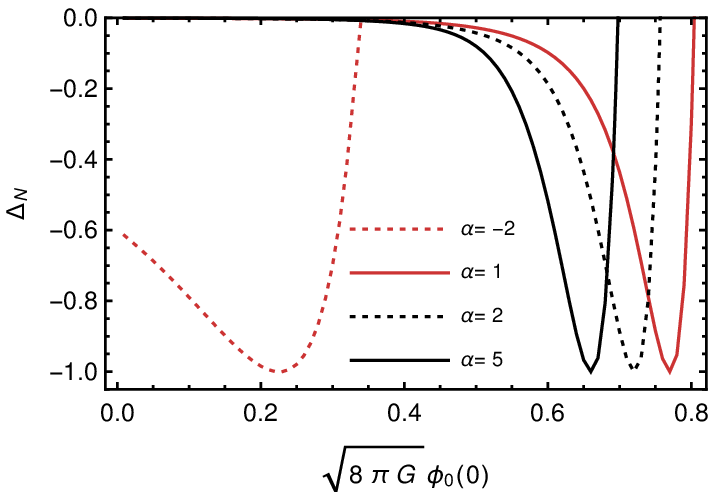, width=0.45 \linewidth}&
\epsfig{file=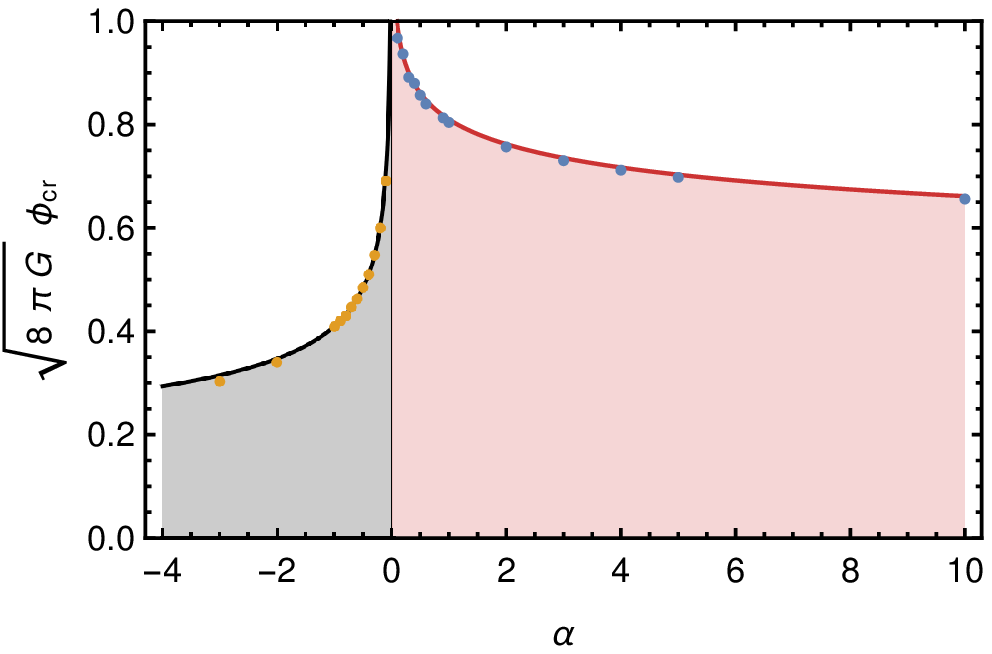, width=0.45 \linewidth}
\end{tabular}
\end{center}
\caption{\label{fig:pCr}Left:Normalized discriminant $\Delta_N=\frac{\Delta}{Max(\left|\Delta\right|)}$ plotted for different coupling constants. Since the physical solutions exist only for negative discriminant, the curve ends $\Delta$ gets closer to zero at $\phi_0(0)=\phi_{cr}$.\\ Right: $\phi_{cr}$ plotted as function of coupling constant. Points are the actual data, while lines are fitted curves given in \ref{eq:fit}. Since the solutions exist only for $\phi_0(0)<\phi_{cr}$, all the boson-stars in sGB gravity lie below these lines}
\end{figure*}
As discussed in \ref{sssec:constraints}, the boson star solutions in sGB gravity exist for $\phi_0(0) \leq \phi_{cr}$, so
that the discriminant \ref{eq:discriminant}, $\Delta \leq 0$.  
This is further illustrated in Figure \ref{fig:pCr} where we present the behavior of the discriminant (rescaled to minimum value -1).
The value of the discriminant approaches zero as $\phi_0(0)$ approaches the critical value $\phi_{cr}$. 
It's not possible to construct solutions with scalar field greater than this value at origin. 
In other words, domain over which solutions can be constructed is now smaller in sGB gravity. Similar bounds on parameter space exist for other boson star solutions such as EDGB fluid stars which exist only up to a maximum central density\cite{Pani}, 5D EGB boson stars 
where we have bounds on the free parameter, $\phi_0(0)$ or $\phi'_0(0)$ \cite{SKV,MBS5D}\\
 Calculating the exact $\phi_{\text{cr}}$ while simultaneously shooting for $\omega$ is quite tedious. We present some 
 approximate values of various physical quantities of boson star in Table \ref{tab:data} upto an error in the last digit. 
 Upon data fitting (Figure \ref{fig:pCr}), we find that
\be
\sigma_{\text{cr}}=\begin{cases}
0.41 \left|\alpha\right|^{-0.24} & \alpha<0\\
0.81 \alpha^{-0.088} & \alpha>0
\end{cases}\label{eq:fit}
\ee
where $\sigma= \phi_0 \sqrt{8\pi G}$. 
This behavior is also displayed in Fig \ref{fig:pCr} where dots are the approximate $\phi_{cr}$ while the lines are fitted curves.
It is clear that $\sigma_{\text{cr}}$ decays faster for negative Gauss Bonnet parameter. As a result, we don't get much room to construct and analyse the solutions in such case. For this reason we'll mainly focus on positive coupling constants.  Similar kind of power law behavior was observed in \cite{MBS5D} for rotating boson stars in 5D EGB gravity ($\alpha_\text{cr} \approx 0.667 \phi'_0(0)^{-2} $, derivative of scalar field at origin ($\phi'_0(0)$) serves as the free parameter and $\phi_0(0)=0$).

\subsection{Binding Energy and Stability}

\begin{figure*}[t]
\begin{center}
\begin{tabular}{lr}
\epsfig{file=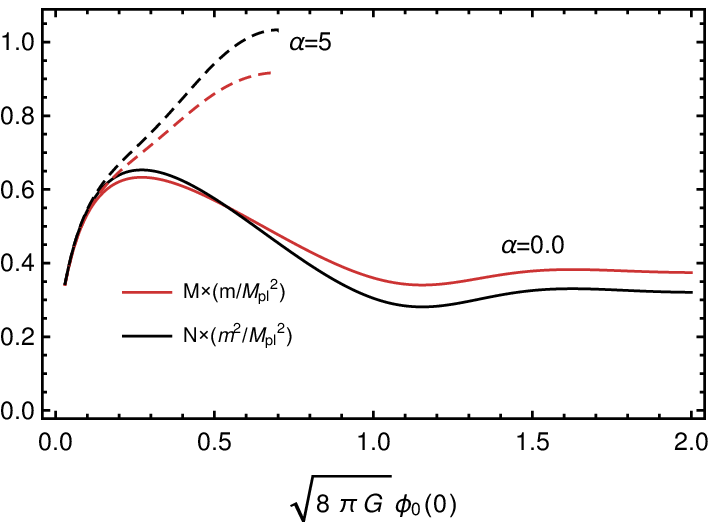, width=0.45 \linewidth}&
\epsfig{file=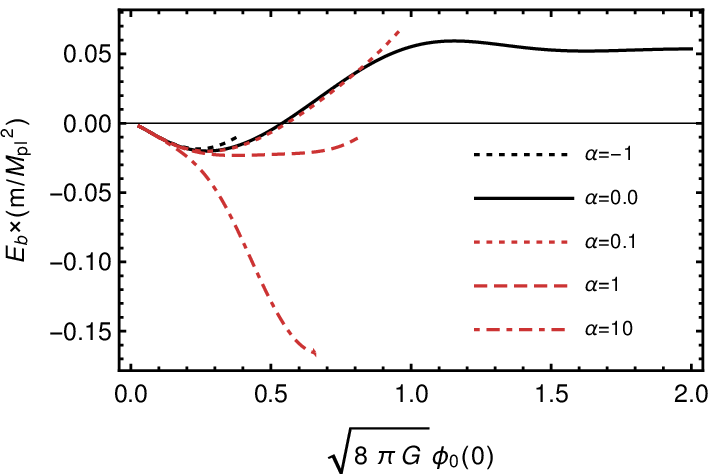, width=0.45 \linewidth}
\end{tabular}
\end{center}
\caption{\label{fig:be}Left: Mass and number of particles for boson stars in Einstein-Hilbert gravity and sGB gravity. Note that in Einstein's gravity, for $\alpha=0$, there's a transition from $N>M$ to $M>N$. On other-hand for $\alpha=5$, N always lies above mass curve. In other words, binding energy is always negative.\\
 Right: Binding energies for different values of coupling constant is displayed. For $\alpha<-0.4$ and $\alpha>0.8$, binding energy is always negative, the configurations can only decay by gravitational collapse. 
}
\end{figure*}
For an  usual boson star without the Gauss-Bonnet interaction, binding energy changes from negative  to positive value
as we increase $\phi_0(0)$. sGB boson stars with small coupling parameter follow the same trend. For positive energies, kinetic energy of particles is greater than the gravitational potential energy holding them down. This causes particles to disperse to infinity. Interestingly for $\alpha>0.800$ and $\alpha<-0.400$, there's peculiar change when all possible values of $\phi_0(0)$ give negative binding energies. These configurations are immune to dispersion, and the only possible mechanism for the instability is gravitational collapse.
The author in reference \cite{bij} also discovered this behavior for boson-star with non minimal coupling to scalar curvature. With coupling of type $\xi\phi \phi^* R$,  they found that $\xi>4$ gives only negative binding energies. Note that we use the similar coupling to Gauss Bonnet term of type   $A \phi \phi^* \mathcal{G}$.\\
More importantly, this occurrence of "classically stable" boson-stars in sGB gravity is in stark contrast with EGB boson-stars in 5 dimensions \cite{MBS5D} that are always classically-unstable. However, having negative binding energy does not guarantee the stability against gravitational collapse. Some configurations might still collapse into Black Holes. Since stability theorems applicable to Einstein-Hilbert boson stars or fluid stars are no longer valid in the present context due to non-minimal gravitational coupling, a complete stability analysis is required for determining gravitational stability. We defer this to a future work.
\begin{figure*}[t]
\begin{center}
\begin{tabular}{lr}
\epsfig{file=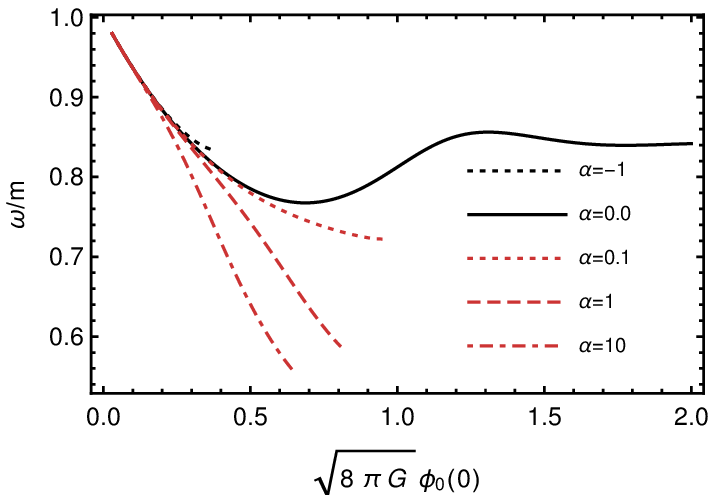, width=0.45 \linewidth}&
\epsfig{file=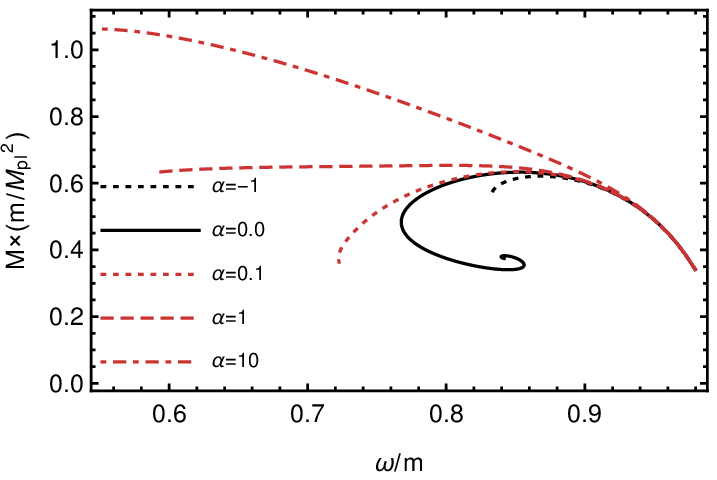, width=0.45 \linewidth}
\end{tabular}
\end{center}
\caption{\label{fig:wCurves}Left: Eigenfrequency $\omega$ found by shooting,  for different values of coupling constants and scalar field at origin,\\ Right: Mass-frequency diagrams in sGB gravity. Although, like 5D EGB boson-stars there is disappearance of the inspiral there aren't any new branches encountered in \cite{MBS5D,BH}
}
\end{figure*}
\subsection{Mass frequency curves}
\begin{figure*}[t]
\begin{center}
\begin{tabular}{lr}
\epsfig{file=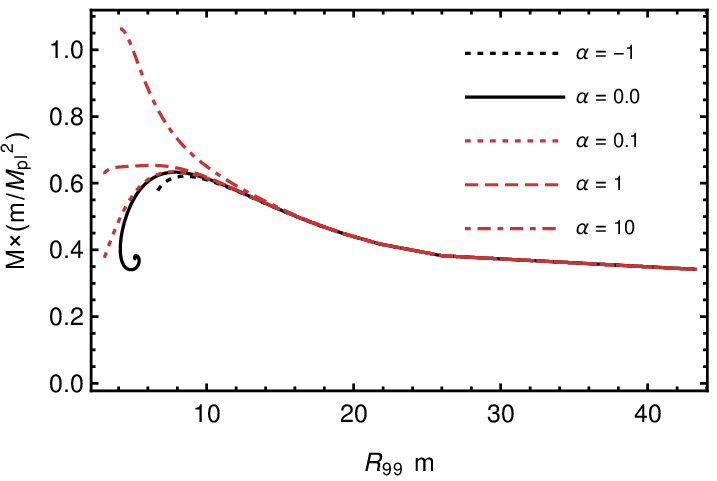, width=0.45 \linewidth}&
\epsfig{file=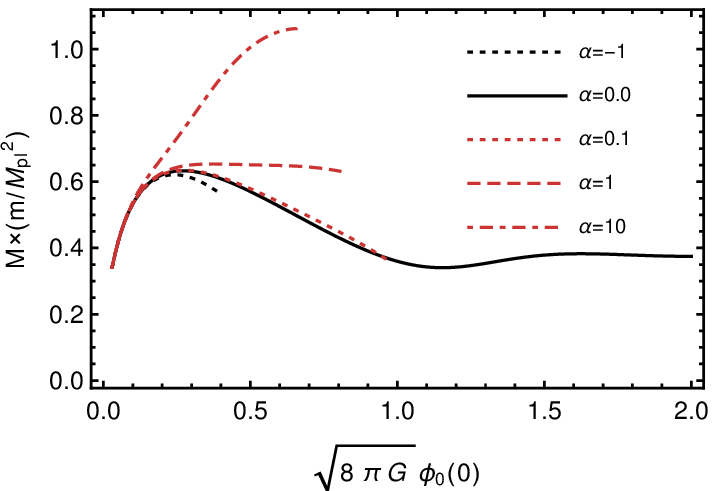, width=0.45 \linewidth}
\end{tabular}
\end{center}
\caption{\label{fig:mCurves}Left: Mass versus radius curves for several values of coupling. Inspiral found in BSs for Einstein gravity ($\alpha=0$), disappears in sGB gravity,\\ Right: Mass as function of the scalar field value at origin. It is shown that the mass increases with the coupling, and for large enough value of coupling are on-to-one function of scalar field.
}
\end{figure*}

 In \cite{BH,MBS5D}, authors have studied different properties of 5D boson 
stars in Gauss Bonnet gravity with frequency $\omega$ as a parameter. They observed spiraling behavior of M-$\omega$ (and N-$\omega$) curves. Depending upon the values of five dimensional Gauss-Bonnet parameter this spiral unwinds giving rise to new branches. See \cite{BH} for more details on this behavior in five dimensions.\\ 
However, for boson stars in sGB gravity, even though the spirals disappear, there are no new branches. This behavior can be better understood by looking at behavior of frequency parameter. As shown in Fig \ref{fig:wCurves} for usual $\alpha =0$ boson star,
frequency $\omega$ is not an injective ( one-to-one) function of free parameter, $\phi_0(0)$. Hence when frequency is itself used as a parameter, multiple masses exist for a given $\omega$. This leads to spiraling behavior. However as we increase the coupling, $\omega$ as function $\phi_0(0)$ takes a value only once and is now injective. Additionally, for large enough coupling, mass is also uniquely defined by the central value of scalar field. (See Fig \ref{fig:mCurves} and Sec.\ref{sssec:MRCurves}). This causes the spiral to disappear.
\subsection{M-R curves \label{sssec:MRCurves}}
One of the reasons for wide-spread popularity of the boson stars, is their strong resemblance to the compact objects like neutron stars. More specifically, 
boson stars and neutron stars exhibit somewhat identical mass versus radius curves \cite{MR}. This makes it worthwhile to construct and study these diagrams in different theories of gravity. As mentioned before, these physical behaviors could
also be very important to constrain the effective gravity theory itself. However, like mass-frequency curve,
the M-R curves for the usual boson star with $\alpha=0$ also has spiraling behavior because of multiple solutions for
the mass (M) for a given value of the radius (R). 
As one increases the parameter $\phi_0(0)$, M-R curves progress into the spiraling region. However, for an sGB boson star, it is no longer possible to have arbitrarily large values of scalar field at origin. This causes inspirals to disappear. Also for sufficiently large coupling, there is one-to-one correspondence between mass and radius i.e both mass and radius are uniquely defined (for eg. $\alpha=10$ in Fig \ref{fig:mCurves}). This occurs for those solutions whose maximum mass is given at $\phi_0(0)=\phi_{cr}$.
Boson stars in 5D Einstein-Gauss-Bonnet theory also shows similar behavior \cite{BH}. 
However M-R curves of sGB boson stars differ from  those of neutron stars found in EDGB gravity theory.  The mass-radius curves of neutron stars in EDGB gravity lie below the curves given by standard Einstein gravity. This is in stark contrast with M-R curves of sGB boson-stars with positive coupling which lie above curve given by $\alpha=0$ (See Figure 2 of \cite{Pani} and compare with positive $\alpha$ in Fig \ref{fig:mCurves}). This peculiar dissimilarity can be attributed to the fact that neutron stars are less massive in EDGB compared to Einstein gravity. Again, it's necessary to remind the reader that the case presented by \cite{Pani} is entirely different fromt the current model, mainly because of the nature of scalar field (Dilaton is a real scalar; also \cite{Pani} do not couple matter field to Gauss-Bonnet term).
\begin{figure*}[t]
\begin{center}
\begin{tabular}{lr}
\epsfig{file=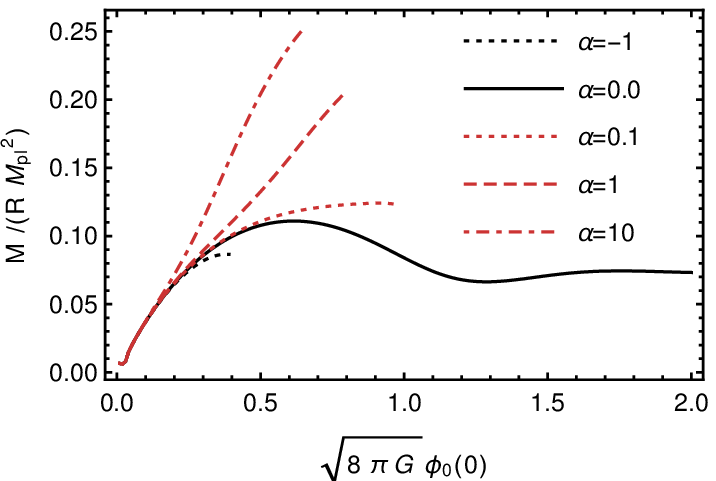, width=0.45 \linewidth}&
\epsfig{file=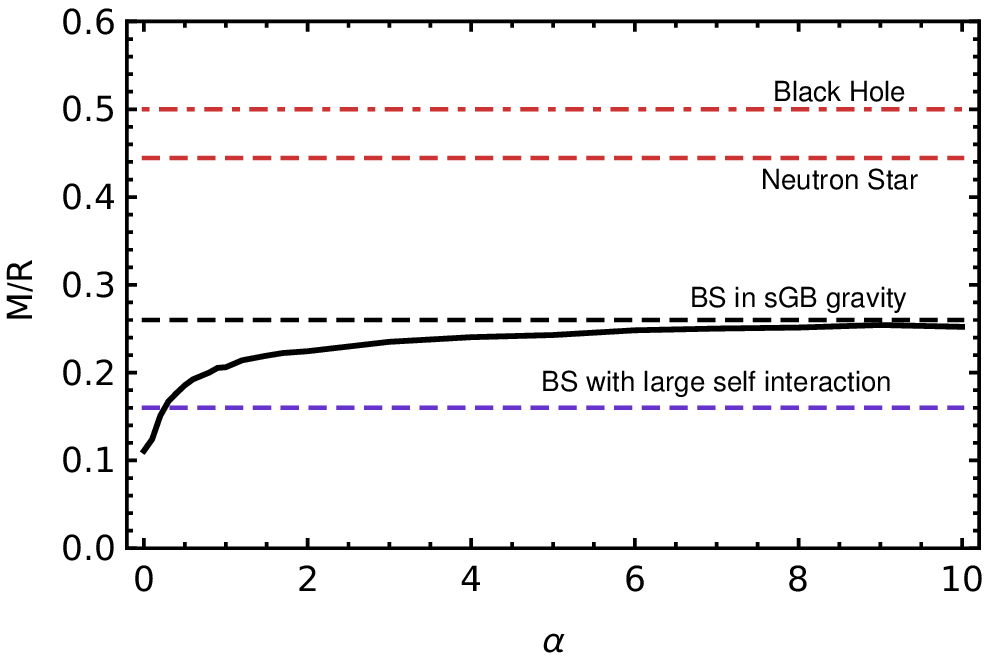, width=0.45 \linewidth}
\end{tabular}
\end{center}
\caption{\label{fig:compact}Left: Compactness as a function of central value of scalar field (with $G=1$),\\ Right: Taking inspiration from \cite{compactness} we plot compactness as a function of coupling. Maximum compactness of self-interacting BS, neutron stars and black holes are displayed for comparison.
}
\end{figure*}
\subsection{Compactness}
As has been explained earlier, compactness of a boson star is a measure of the amount of mass contained
within the effective radius of the same. For example, compactness of a Schwarzschild black hole is $C_{BH}=1/2$,
while the maximum compactness of a non-rotating fluid stars $C_\text{Max,NS}=4/9$ is given by Buchdahl limit \cite{buchdahl}. 
The compactness of the stable self-interacting boson stars was studied in \cite{compactness} for Einstein Gravity, where they found
an upper bound on the ratio M/R to be 0.16 for large self interactions. We found that sGB boson stars are much more compact than the regular self interacting ones and more than half as compact as the Black Holes.
In sGB gravity, ADM mass of a boson star increases with the Gauss Bonnet coupling, with a negligible decrease in the effective radius $R_{99}$. This fact essentially increases the compactness of the sGB boson stars. Figure \ref{fig:compact} displays the compactness of sGB boson stars as a function of coupling parameter $\alpha$. As shown, compactness increases with $\alpha$, reaching an asymptotic value of $C_\text{Max,sGB-BS}\approx 0.253$. Compactness of black hole $C_{BH}=1/2$ and maximum compactness of neutron stars  $C_\text{Max,NS}=4/9$ and self-interacting BS $C_\text{Max,BS}=0.16$ are also displayed for comparison. 

\section{Summary and Future Work}
In this paper, we have constructed the boson stars solution in scalar-Gauss-Bonnet gravity in four dimension. 
The action contains a Gauss-Bonnet term coupled to the square of scalar field amplitude. Our results show that the scalar field can
not have arbitrarily high values at the center of the boson star. This feature is shared by stars in other theories of modified gravity as well (\cite{Maselli:2016gxk,BH,Pani}).  This can be better understood by looking at the behavior of scalar field near the center of a star.
This imposes an upper bound on the value of the scalar field at the origin, $\phi_0(0)$. \\
We have studied various interesting properties of boson stars specifically focusing on their
dependence on Gauss-Bonnet coupling parameter $\alpha$. 
We studied mass-frequency (M-$\omega$) curves, which no longer have the spiraling feature found in boson stars without scalar-Gauss-Bonnet coupling. This disappearance of the inspiral can be observed
in 5D Gauss-Bonnet theory as well, even though shape of curve can be qualitatively different from M-$\omega$ diagrams of sGB boson stars.  Spirals found in Mass-Radius diagrams disappear as well and for large enough coupling, mass and radius are quite unique, i.e for a given mass, boson-star has an unique radius. Nevertheless M-R curves maintain an appearance similar to those of Neutron stars \cite{MR} and boson-stars in 5D EGB gravity. 
Furthermore, boson star with positive coupling are more massive and due to a slight decrease in radius, they are also more compact.
However in our analysis, we have not included the self-interaction of the scalar field. The resulting boson stars have masses (of order $M_{pl}^2/m$) much smaller than the Chandrasekhar mass of fermionic stars  (of order $M_{pl}^3/m^2$). To reach astrophysical mass scales, a repulsive self interaction term is needed. Since in sGB gravity, there is a very slight decrease in stellar radius, we believe that inclusion of self interaction can create very compact stars.\\
 A complete stability analysis is beyond the scope of this paper,and is left for our future work. Nonetheless, we have made some comments about classical stability of our boson star configurations. Interestingly for $\alpha<-0.4$ and $\alpha>0.8$, binding energy is always negative. This implies that these stars are immune to decay by dispersion, though they can still collapse into black-holes. This is very different from 5D EGB boson stars which are always classically unstable \cite{MBS5D}.

\bibliographystyle{h-physrev4}

\begin{thebibliography}{99}
\bibitem{Liebling:2012fv} 
  S.~L.~Liebling and C.~Palenzuela,
  Living Rev.\ Rel.\  {\bf 15}, 6 (2012)
  doi:10.12942/lrr-2012-6
  [arXiv:1202.5809 [gr-qc]].
  
\bibitem{MBS5D} 
  Y.~Brihaye and B.~Hartmann,
  Class.\ Quant.\ Grav.\  {\bf 33}, no. 6, 065002 (2016)
  doi:10.1088/0264-9381/33/6/065002
  [arXiv:1509.04534 [hep-th]].
  
\bibitem{SKV} 
  L.~J.~Henderson, R.~B.~Mann and S.~Stotyn,
  Phys.\ Rev.\ D {\bf 91}, no. 2, 024009 (2015)
  doi:10.1103/PhysRevD.91.024009
  [arXiv:1403.1865 [gr-qc]].
  
\bibitem{RBS5D} 
  Y.~Brihaye and J.~Riedel,
  Phys.\ Rev.\ D {\bf 89}, no. 10, 104060 (2014)
  doi:10.1103/PhysRevD.89.104060
  [arXiv:1310.7223 [gr-qc]].
  
  
\bibitem{BH} 
  B.~Hartmann, J.~Riedel and R.~Suciu,
  Phys.\ Lett.\ B {\bf 726}, 906 (2013)
  doi:10.1016/j.physletb.2013.09.050
  [arXiv:1308.3391 [gr-qc]].
  
\bibitem{Gleiser:1988rq} 
  M.~Gleiser,
  Phys.\ Rev.\ D {\bf 38}, 2376 (1988)
  Erratum: [Phys.\ Rev.\ D {\bf 39}, no. 4, 1257 (1989)].
  doi:10.1103/PhysRevD.38.2376, 10.1103/PhysRevD.39.1257
  
\bibitem{bsnm} 
  A.~Marunovic,
  arXiv:1512.05718 [gr-qc].
  
\bibitem{bij} 
  J.~J.~van der Bij and M.~Gleiser,
  Phys.\ Lett.\ B {\bf 194}, 482 (1987).
  doi:10.1016/0370-2693(87)90221-8
  
\bibitem{DE} 
  L.~N.~Granda,
  Mod.\ Phys.\ Lett.\ A {\bf 27}, 1250018 (2012)
  doi:10.1142/S0217732312500186
  [arXiv:1108.6236 [hep-th]].
  
\bibitem{Pani} 
  P.~Pani, E.~Berti, V.~Cardoso and J.~Read,
  Phys.\ Rev.\ D {\bf 84}, 104035 (2011)
  doi:10.1103/PhysRevD.84.104035
  [arXiv:1109.0928 [gr-qc]].
  
\bibitem{compactness} 
  P.~Amaro-Seoane, J.~Barranco, A.~Bernal and L.~Rezzolla,
  JCAP {\bf 1011}, 002 (2010)
  doi:10.1088/1475-7516/2010/11/002
  [arXiv:1009.0019 [astro-ph.CO]].
  
\bibitem{MR} 
  G.~A.~Carvalho, R.~M.~M.~Jr and M.~Malheiro,
  J.\ Phys.\ Conf.\ Ser.\  {\bf 630}, no. 1, 012058 (2015).
  doi:10.1088/1742-6596/630/1/012058
  
\bibitem{buchdahl} 
  H.~A.~Buchdahl,
  Phys.\ Rev.\  {\bf 116}, 1027 (1959).
  doi:10.1103/PhysRev.116.1027
  
  \bibitem{Maselli:2016gxk} 
  A.~Maselli, H.~O.~Silva, M.~Minamitsuji and E.~Berti,
  Phys.\ Rev.\ D {\bf 93}, no. 12, 124056 (2016)
  doi:10.1103/PhysRevD.93.124056
  [arXiv:1603.04876 [gr-qc]].
  
\bibitem{Jetzer:1991jr} 
  P.~Jetzer,
  Phys.\ Rept.\  {\bf 220}, 163 (1992).
  doi:10.1016/0370-1573(92)90123-H
  
\bibitem{Nojiri:2007te} 
  S.~Nojiri, S.~D.~Odintsov and P.~V.~Tretyakov,
  Phys.\ Lett.\ B {\bf 651}, 224 (2007)
  doi:10.1016/j.physletb.2007.06.029
  [arXiv:0704.2520 [hep-th]].
  
\bibitem{Astashenok:2014nua} 
  A.~V.~Astashenok, S.~Capozziello and S.~D.~Odintsov,
  JCAP {\bf 1501}, no. 01, 001 (2015)
  doi:10.1088/1475-7516/2015/01/001
  [arXiv:1408.3856 [gr-qc]].
  
\end{thebibliography}

\end{document}